\documentclass{article}

    \usepackage[preprint]{neurips_2020}

\usepackage[utf8]{inputenc} 
\usepackage[T1]{fontenc}    
\usepackage{hyperref}       
\usepackage{url}            
\usepackage{booktabs}       
\usepackage{amsfonts}       
\usepackage{amsmath}        
\usepackage{nicefrac}       
\usepackage{microtype}      
\usepackage{graphicx}
\usepackage{grffile}
\usepackage{bmpsize}
\usepackage{svg}
\usepackage{makecell}
\usepackage{bbm}
\usepackage{array}

\title{CO-Search: COVID-19 Information Retrieval with Semantic Search, Question Answering, and Abstractive Summarization}

\author{%
  Andre Esteva\\
  Salesforce Research\\
  Palo Alto, CA 94301 \\
  \texttt{aesteva@salesforce.com} \\
  \And
  Anuprit Kale \\
  Salesforce Research\\
  Palo Alto, CA 94301 \\
  \texttt{akale@salesforce.com} \\
  \And
  Romain Paulus \\
  Salesforce Research\\
  Palo Alto, CA 94301 \\
  \texttt{rpaulus@salesforce.com} \\
  \And
  Kazuma Hashimoto \\
  Salesforce Research\\
  Palo Alto, CA 94301 \\
  \texttt{k.hashimoto@salesforce.com} \\
  \And
  Wenpeng Yin  \\
  Salesforce Research\\
  Palo Alto, CA 94301\\
  \texttt{wyin@salesforce.com} \\
  \And
  Dragomir Radev  \\
  Salesforce Research\\
  Palo Alto, CA 94301\\
  \texttt{dradev@salesforce.com} \\
  \And
  Richard Socher  \\
  Salesforce Research\\
  Palo Alto, CA 94301\\
  \texttt{rsocher@salesforce.com} \\
}

\begin{document}

\maketitle

\begin{abstract}
The COVID-19 global pandemic has resulted in international efforts to understand, track, and mitigate the disease, yielding a significant corpus of COVID-19 and SARS-CoV-2-related publications across scientific disciplines.  As of May 2020, 128,000 coronavirus-related publications have been collected through the COVID-19 Open Research Dataset Challenge \cite{wang2020cord}. 
Here we present CO-Search, a retriever-ranker semantic search engine designed to handle complex queries over the COVID-19 literature, potentially aiding overburdened health workers in finding scientific answers during a time of crisis. 
The retriever is built from a Siamese-BERT\cite{reimers2019sentence} encoder that is linearly composed with a TF-IDF vectorizer \cite{shahmirzadi2019text}, and reciprocal-rank fused \cite{cormack2009reciprocal} with a BM25 vectorizer. 
The ranker is composed of a multi-hop question-answering module\cite{Asai&al.2020}, that together with a multi-paragraph abstractive summarizer adjust retriever scores. 
To account for the domain-specific and relatively limited dataset, we generate a bipartite graph of document paragraphs and citations, creating 1.3 million (citation title, paragraph) tuples for training the encoder. 
We evaluate our system on the data of the TREC-COVID\cite{voorhees2020trec} information retrieval challenge. 
CO-Search obtains top performance on the datasets of the first and second rounds, across several key metrics:  normalized discounted cumulative gain, precision, mean average precision, and binary preference. 
\end{abstract}

\section{Introduction}

The global response to COVID-19 has yielded a growing corpus of scientific publications - increasing at a rate of thousands per week - about COVID-19, SARS-CoV-2, other coronaviruses, and related topics. The individuals on the front lines of the fight - healthcare practitioners, policy makers, medical researchers, etc. - will require specialized tools to keep up with the literature.

CO-Search is a retriever-ranker semantic search engine that takes search queries (including questions in natural language), and retrieves scientific articles over the coronavirus literature. 
CO-Search displays content from over 128,000 coronavirus-related scientific papers made available through the COVID-19 Open Research Dataset Challenge (CORD-19) \cite{wang2020cord} - an initiative put forth by the US White House and other prominent institutions in early 2020.

Retrieval is done using a semantic model, and two keyword models. For the semantic model, we create a bipartite graph from paragraphs and their cited articles, to generate over 2.2 million (paragraph, title) tuples that we use to train a Siamese-BERT (SBERT) model \cite{reimers2019sentence} on the binary task of classifying a title as being cited by a paragraph. SBERT \cite{devlin2018bert} is used to embed queries and documents into the same latent space, enabling nearest-neighbor semantic retrieval.
Further, we combine these embeddings with TFIDF and BM25. Specifically, we linearly combine SBERT paragraph-level retrieval scores with TF-IDF document-level retrieval scores to generate a document list, then use reciprocal ranked fusion to combine this list with that obtained from BM25 retrieval using Anserini \cite{yang2017anserini}.

Ranking takes the retrieved scores and modulates them with a question answering module, and an abstractive summarizer. We train a multi-hop question answering model with Wikipedia, following \cite{Asai&al.2020}, that treats the query as a question and generates answers from the retrieved documents. We train an abstractive summarizer (composed of a BERT encoder and modified GPT-2 decoder \cite{radford2019language}) in the same way to generate a summary. 
Our ranker then takes the scores of the retrieved documents and modulates them based on the degree to which the documents contain the generated answers and summaries, to return a ranked document set.   

We evaluate CO-Search on data from the first two rounds of the TREC-COVID challenge \cite{voorhees2020trec} - a five-round IR competition for COVID-19 search engines - using several standard metrics: normalized discounted cumulative gain (nDCG), precision with N documents (P@N), mean average precision (MAP), and binary preference (Bpref). 
TREC-COVID considers IR system submissions that are either \emph{manual} - in which queries and retrieved documents may be manually adjusted by a human operator - or \emph{automatic} - in which they may not.
A third category is accepted in Rounds 2-5, of type \emph{feedback}, in which systems are trained with supervision from the annotations of prior rounds.
Submissions compete on a pre-defined set of topics, and are judged using various metrics, including those listed above.
Expert human annotators provide relevance judgements on a small set of topic-document pairs, which are included, together with non-annotated pairs, in the evaluation.

CO-Search is an automated system. On Round 1 and Round 2 data, evaluated against other automated systems using strictly judged topic-document pairs, it achieves a rank of \#1 on nDCG@10, P@5, P@10, MAP, and Bpref (full details in Section ~\ref{section:evaluation}). Taken one step further by evaluating against all systems and all topic-documents (judged \& non-judged), CO-Search ranks in the top 21 on Round 1, and top 3 on Round 2.

We do not propose a novel algorithm, but rather, we present a functioning system and open-source code, rigorously evaluated, to support the global effort and encourage others to build on our work.

\section{Related Work}
The CORD-19 Challenge's \cite{wang2020cord} coronavirus-related corpus, primarily from PubMed, mostly published in 2020, has quickly generated a number of data science and computing works \cite{Bullock&al.2020}. These cover topics from information retrieval to natural language processing, including applications in question answering \cite{Tang&al.2020}, text summarization,  and document search \cite{voorhees2020trec}.

\paragraph{Information Retrieval.} Over the last three months, more than 20 organizations have launched publicly accessible search engines using the CORD-19 corpus. For instance, Neural Covidex \cite{Zhang&al.2020} was constructed from various open source IR building blocks, including Pyserini, Blacklight, and Apache Solr, for retrieval, as well as a T5 transformer \cite{raffel2019exploring}, finetuned on the MS MARCO dataset \cite{bajaj2016ms}, to predict query-document relevance, for ranking. SLEDGE \cite{MacAvaney&al.2020} extends this by using SciBERT \cite{Beltagy&al.2019}, also fine-tuned on MS MARCO, to re-rank articles retrieved with BM25.

\paragraph{Question Answering and Text Summarization.} One of the first question answering systems built on top of the CORD-19 corpus is CovidQA\footnote{http://covidqa.ai}\cite{Tang&al.2020}, which includes a small number of questions from the CORD-19 tasks. A multi-document summarization system, CAiRE, is described in \cite{Su&al.2020}. It generates abstractive summaries using a combination of UniLM\cite{Dong&al.2019} and BART\cite{Lewis&al.2019}, fine tuned on a biomedical review data set.




\section{Dataset}
\label{section:dataset}
To evaluate our system, we combine the CORD-19 corpus with the the TREC-COVID competition’s evaluation dataset: topics, and relevance judgements which annotate query-document pairs into either irrelevant, partially relevant, or relevant.

\subsection{Documents}
The U.S. White House, along with the U.S. National Institutes of Health, the Allen Institute for AI, the Chan-Zuckerberg Initiative, Microsoft Research, and Georgetown University recently prepared the CORD-19 Challenge in response to the global crisis. 
As of May 2020, this resource consists of over 128,000 scientific publications (up from 29,000 at the challenge inception in February 2020) about COVID-19, SARS-CoV-2, and earlier coronaviruses \cite{wang2020cord}.

This challenge represents a call-to-action to the AI and IR communities to “develop text and data mining tools that can help the medical community develop answers to high priority scientific questions”. 
It is currently the most extensive coronavirus literature corpus publicly available. 

The tasks put forth by the CORD-19 challenge are broad, open-ended, and qualitative, with submissions scored by human experts based on accuracy, documentation, and presentation. 
The unstructured nature of the challenge has resulted in submissions ranging from IR systems to interactive visualizations. 
As such, we do not use these tasks to evaluate our system.

\subsection{Evaluation Topics}

\begin{table}
    \caption{Sample TREC COVID topic from \cite{voorhees2020trec}}
    \centering
    \begin{tabular}{p{2cm}|p{10cm}}
        \toprule
        \textbf{query} & coronavirus drug repurposing  \\
        \midrule
        \textbf{question} & which SARS-CoV-2 proteins-human proteins interactions indicate potential for drug 
                            targets.
                            Are there approved drugs that can be re-purposed based on this information? \\
        \midrule
        \textbf{narrative} & Seeking information about protein-protein interactions for any of the SARS-CoV-2 structural proteins that represent a promising therapeutic target, and the drug molecules that may inhibit the virus and the host cell receptors at entry step. \\
        \bottomrule
    \end{tabular}
    \label{tab:TREC-COVID_topic}
\end{table}





In response to CORD-19, the Text Retrieval Conference (TREC) recently partnered with the National Institute of Standards and Technology (NIST), to define a structured and quantitative evaluation system for coronavirus IR systems. 
The TREC-COVID challenge \cite{voorhees2020trec} is composed of 5 successive rounds of evaluation on 30-50 topics.  
The first round includes 30 topics. Each subsequent round takes the prior round’s topics and adds five new ones.

Each topic is represented as a tuple consisting of a query, a question, and a narrative, with an increasing amount of detail in each (see Table~\ref{tab:TREC-COVID_topic}). 
IR systems must retrieve up to 1,000 ranked documents per topic from the CORD-19 publications, and are evaluated on a set of metrics including nDCG@10, p@N, MAP, and Bpref.

\section{System Architecture}

\begin{figure}
    \centering
    \includegraphics[width=\columnwidth]{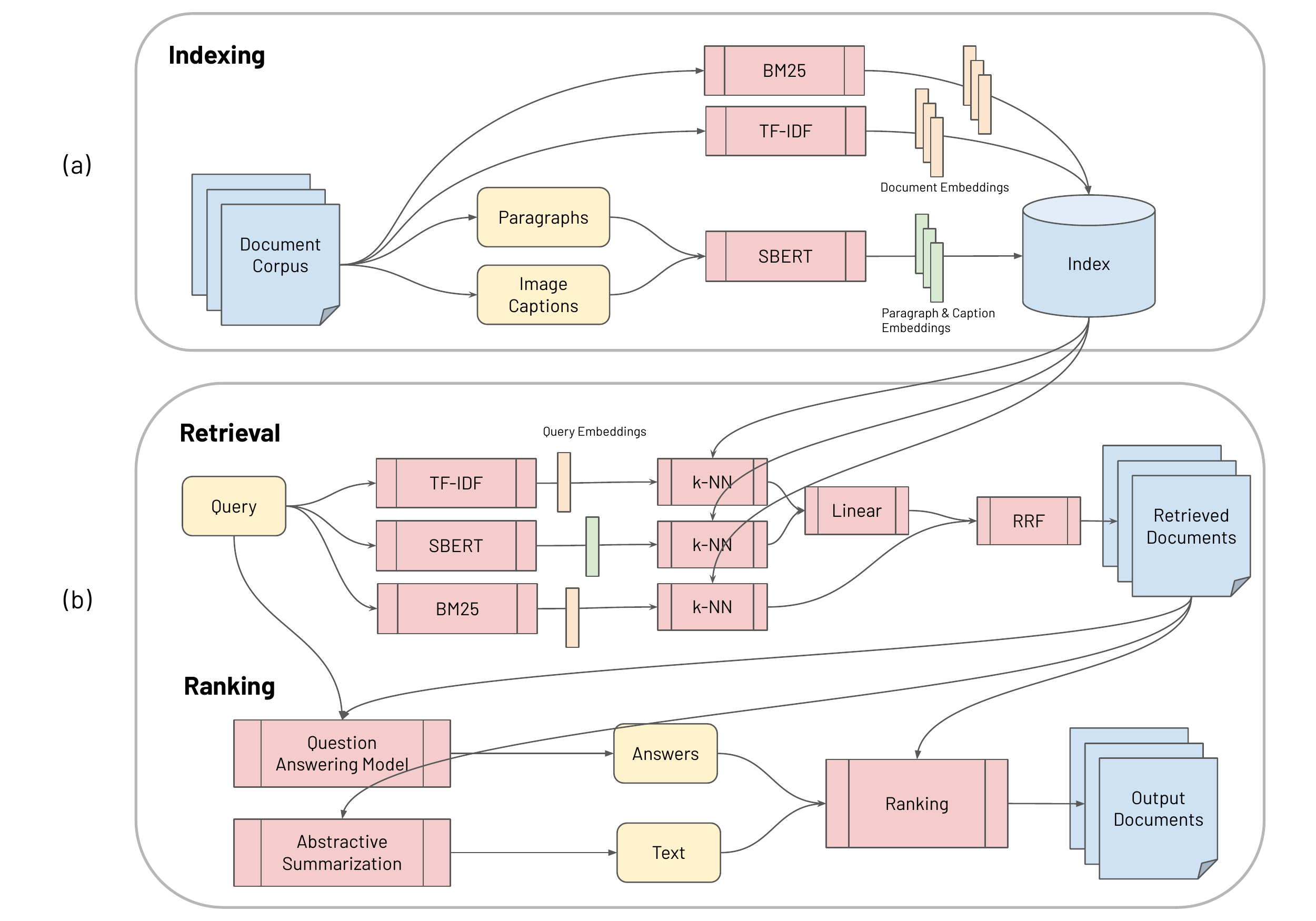}
    \caption{\textbf{System architecture.} (a) Documents are split into paragraphs and image captions, then embedded with a pretrained Siamese BERT network and stored into an index, along with the TF-IDF \& BM25 vectors of the entire documents. (b) The system computes a linear combination of TF-IDF and SBERT retrieval scores, then combines them via reciprocal ranked fusion with the retrieval scores of BM25. The retrieved documents and query are parsed through a question answering model and an abstractive summarizer prior to being ranked based on answer match, summarization match, and retrieval scores.}
    \label{fig:system_architecture}
\end{figure}

CO-Search consists of a retriever and a ranker, with an offline pre-processing step to create a document index.
The index is created by embedding documents in three ways: a pre-trained SBERT model embeds paragraphs, and both a TF-IDF vectorizer and BM25 vectorizer (using the Anserini framework \cite{yang2017anserini}). See Fig. \ref{fig:system_architecture}(a).
Retrieval is done by linearly composing query-paragraph similarity scores of the SBERT model with query-document scores from TF-IDF, then fusing them via reciprocal rank fusion \cite{cormack2009reciprocal} with query-document scores from the BM25 model. 
This defines a retrieval score for each document, given a query.
Ranking takes this set of documents, runs them through a question answering module (QA) and an abstractive summarizer, then ranks the documents by a weighted combination of their retrieval scores, the QA output, and the summarizer output. Full details below.

\subsection{Indexing}
\label{section:indexing}
We use a semantic method (BERT embeddings) to embed paragraphs and image captions, and two keyword methods (TFIDF, BM25) to embed entire documents. See Fig. \ref{fig:system_architecture}(a).

\begin{figure}
    \centering
    \includegraphics[width=\columnwidth]{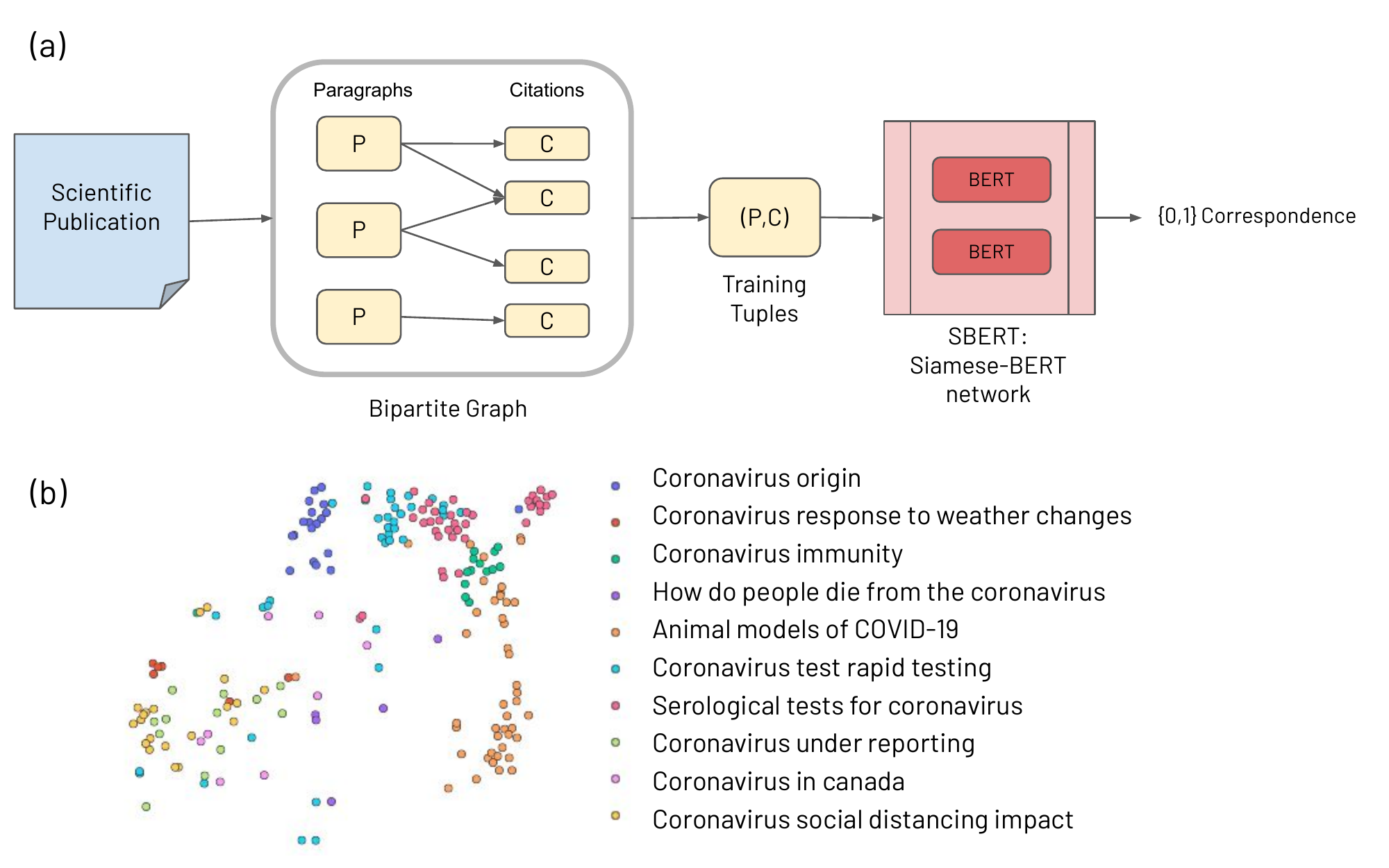}
    \caption{\textbf{Semantic embeddings.} (a) Documents are split into paragraphs and their contained citations to form a bipartite graph that induces training tuples $(p,c)$. These are fed to a Siamese-BERT (SBERT) model trained to discern if a citation is contained in a given paragraph. (b) t-SNE visualization of the SBERT embeddings of paragraphs, colored by relevance to 10 SBERT-embedded topics.}
    \label{fig:sbert}
\end{figure}

In this use-case, semantic embeddings face the challenge of working with a relatively small number of long documents.
To address this, we split the documents into paragraphs, extract the titles of the citations of each paragraph, and form a bipartite graph of paragraphs and citations with edges implying that a citation $c$ came from a paragraph $p$. 
We use the graph to form tuples - $(p,c) \text{ s.t. } c \in p$ - for training an SBERT model \cite{reimers2019sentence} to classify if a title was cited by a paragraph.
Additionally, we generate an equivalent number of negative training samples of incorrect tuples - $(p,c) \text{ s.t. } c \notin p$.
We train the model with cross-entropy loss, Adam optimization \cite{kingma2014adam} with a learning rate of $2\mbox{\sc{e}-}5$, a linear learning rate warmup over 10\% of the training data, and a default pooling strategy of MEAN.
See Fig. \ref{fig:sbert}(a).

The structure of the learned latent space embeds queries near documents that share semantic meaning. 
Visualizing this reveals a human-understandable clustering of documents and topics. 
Fig. \ref{fig:sbert}(b) shows a two-dimensional t-SNE \cite{maaten2008visualizing} plot of the embedded space, with different colors representing topics of TREC-COVID, and points representing documents. 
Qualitatively we observe that semantically similar documents cluster by topic. 

\subsection{Retrieval}
At runtime, the retrieval step takes an input query, embeds it using SBERT, computes approximate nearest neighbors over the SBERT paragraph embeddings, and returns a set of paragraphs, together with each paragraph's cosine similarity to the query.
TF-IDF and BM25 operate at the document level, each returning vectors $t \in \mathbb{R}^M$ and $b \in \mathbb{R}^M$ such that $t_i = \text{TFIDF}(\text{query}, \text{document } i)$, $b_i = \text{BM25}(\text{query}, \text{document } i)$, and $M$ is the size of the document corpus.

The SBERT and TF-IDF scores are combined linearly. For document $d$ (containing paragraphs $p$), and query $q$, with subscript $es$ denoting an SBERT embedding, their combination $C$ is given by:
\begin{equation}
    C(q, d) = \mu \max_{p \in d}[\cos(p_{es}, q_{es})] + (1 - \mu) \text{TFIDF}(q, d)
\end{equation}
This induces a ranking $R_C^q$ on the documents, which is then combined with the BM25-induced ranking $R_B^q$ using reciprocal ranked fusion \cite{cormack2009reciprocal}, to obtain a final retrieved ordering:
\begin{equation}
    RRF(q, d) = \frac{1}{k + R_C^q(d)} + \frac{1}{k + R_B^q(d)}
\end{equation}
In practice, we find that the constants $\mu=0.7$ and $k=60$ yield good results. Future work could consider using a learned layer to attend over semantic and keyword embeddings, given the query. 

\subsection{Ranking}
Ranking combines the RRF scores of the retrieved documents with the outputs of the QA engine and the summarizer. 
We define $Q$ to measure the degree to which a document answers a query:
\begin{equation}
Q(q, d) = 1.1 ^ N \text{, with } N = \sum_{a \in A(q)} \mathbbm{1}(a \in d)
\end{equation}
Where $\mathbbm{1}(x)$ is the indicator function: $\mathbbm{1}(x) = \{1$ if $x$ is true, 0 otherwise\}. The set $A(q)$ contains the text span outputs of the QA model. 
We define $S$ to measure the degree to which a document summarizes the set of documents retrieved for a query:
\begin{equation}
S(q, d) = \frac{1}{2} + \frac{1}{2} \max_{p \in d} cos(p_e, M(q)_e) 
\end{equation}

Where $M(q)_e$ is the embedded abstractive summary of $q$, summarized across all retrieved documents. Then the final ranking score $R(d,q)$ of a document, for a particular query, is given by:
\begin{equation}
R(q, d) = S(q, d) \cdot Q(q, d) \cdot RRF(q, d)
\end{equation}
With higher scores indicating better matches.
In essence, rank score $R$ is determined by letting $S$ and $Q$ modulate the retrieval score of a query-document pair.

\subsubsection{Question Answering} 
Whereas standard question answering systems generate answers, our model extracts multiple answer candidates (text spans) from the retrieved paragraphs, which are then used downstream for ranking.

The QA model takes the query, the retrieved paragraphs, and uses a sequential paragraph selector model, following \cite{Asai&al.2020}, to filter for paragraphs that could answer the query. 
Specifically, the model uses multi-hop reasoning to model relationships between paragraphs, and select sequentially ordered sets of them (as opposed to simply providing relevance scores for each paragraph, independently).
It is pre-trained using HotpotQA \cite{yang2018hotpotqa} - a Wikipedia-derived dataset of 113k question-answer pairs and sentence-level supporting facts, and further fine-tuned on the PubMedQA dataset \cite{jin2019pubmedqa} for biomedical specificity. 
Following \cite{Asai&al.2020}, we use paragraphs with high TF-IDF scores for the given query as negative examples for the sequential paragraph selector. 
The original beam search is modified to include paragraph diversity and avoid extracting the same answers from different paths. 

Once filtered, the paragraph sets are fed into an extractive reading comprehension model - trained on the SQuAD \cite{rajpurkar2016squad} dataset, given its similarity to the TREC topics - to extract answer candidates. 

\subsubsection{Abstractive Summarization} 
Our summarizer takes the retrieved documents and generates a single abstractive summary.
It is an encoder-decoder model, with BERT \cite{devlin2018bert} as the encoder and a modified GPT-2 model \cite{radford2019language} as the decoder. 

We extend the original GPT-2 model by adding a cross-attention function alongside every existing self-attention function. 
We constrain the cross-attention function to attend strictly to the final layer outputs of the encoder.
We use the base models of \cite{wolf2019transformers}, with 12 layers, 768-dimensional activations in the hidden layers, and 12 attention heads. 
The model is pre-trained using self-supervision with a gap-sentence generation objective \cite{zhang2019pegasus}, followed by single-document supervised training, using CORD-19 full documents as input, and the paper abstract as target output.

To increase the probability that a generated summary matches (and thus, helps re-rank) the contents of the retrieved paragraphs, we architect the model to generate short summaries, following \cite{fan2017controllable}.
Abstracts are split into five groups based on the number of tokens: <65, 65-124, 125-194, 195-294, >295. 
During training, a special token is provided to specify the summary length in these 5 categories.
At inference time, the model is initialized to output summaries of token lengths <65. 

To adjust the model to operate on multiple retrieved paragraphs, we concatenate the first four sentences of the retrieved paragraphs until they reach an input length of 512 tokens, then feed this into the summarization model.

\section{Evaluation}
\label{section:evaluation}



We evaluate our system quantitatively using the CORD-19 document dataset, and the topics and relevance judgements provided by the TREC-COVID competition (see Section \ref{section:dataset} for full details).

In each of its five rounds, the competition provides a set of topics - as (query, question, narrative) tuples - to evaluate submissions, along with relevance judgements - scores of 0 for irrelevance, 1 for partial relevance, and 2 for relevance - on a small subset of all possible topic-document pairs. 
Teams submit up to 1,000 ranked documents per query, and the organizers pool from amongst the most common topic-document pairs for judging (depth-7 pooling, in which the top 7 documents from each response provided
by the set of contributing systems are judged for relevance by human assessors \cite{lu2016effect}). 
This yields extremely sparse and biased labeling. 
As a result, it is critical to evaluate systems both on the full dataset and on the annotated subsets. 

Given the inherent difficulty of comparing IR systems, TREC has historically used a number of metrics to judge them. 
Key amongst them are high-precision metrics such as nDCG@10 (which the leaderboards use), precision@5, precision@10, and MAP.
The critical limitation with these is that their effectiveness relies on complete relevance judgements across all topic-document pairs, which is intractable. 
To account for this, the competition considers measures that work with incomplete relevance judgements, such as bpref.

\subsection{Metrics}
Below we define key metrics in evaluation. Throughout this work we adopt the standard convention that m@N refers to an evaluation using metric m, and the top N retrieved documents.

\paragraph{Precision (P):}
\begin{equation}
P@N = \frac{|\text{\{relevant documents in top-N\}}|}{N} 
\end{equation}


\paragraph{Normalized discounted cumulative gain (nDCG):} 
For position $i \in \{0, 1, …, N\}$, the normalized discounted cumulative gain of a retrieved s    et of documents over $Q$ queries is given by:




\begin{equation}
\text{nDCG}@N =\frac{1}{Q} \sum_{q=1}^{Q} \frac{\text{DCG}_{p}^{(q)} }{\text{IDCG}_{p}^{(q)}}
\text{, with }
\text{DCG}_{p}^{(q)} = \text{rel}_1^{(q)} + \sum_{i=2}^{N} \frac{\text{rel}_i^{(q)}}{log_2(i)} 
\end{equation}

Where $\text{rel}_i^{(q)}$ denotes the relevance of entry $i$, ranked according to query $q$.
IDCG denotes the ideal and highest possible DCG. 
In the limit of perfect annotations, nDCG performs reliably in measuring search engine performance. 
Since it treats non-annotated documents as incorrect ($\text{rel}_i$ evaluates to zero), it is less reliable for datasets with incomplete annotations. 

\paragraph{Mean average precision (MAP):}
The average precision (AP) of a retrieved document set is defined as the integral over the normalized precision-recall curve of the set’s query. 
MAP is defined as the mean AP over all queries: 
\begin{equation}
\text{MAP}= \frac{1}{Q} \sum_{q=1}^{Q} \int_0^1 P_q(R)dR
\end{equation}

Where R is recall, $P_q$ is precision as a function of recall, for a particular query. 
Note that, as in the case of nDCG, MAP penalizes search engines that yield accurate but unique (i.e. non-annotated) results, since non-annotated documents are treated as irrelevant by $P$. 

\paragraph{Binary preference (Bpref).}
Bpref strictly uses information from judged documents. 
It is a function of how frequently relevant documents are retrieved before non-relevant documents. 
In situations with incomplete relevance judgements (most IR datasets) it is more stable than other metrics, and it is designed to be robust to missing relevance judgements. It gives roughly the same results with incomplete judgements as MAP would give with complete judgements \cite{lu2016effect}. 
It is defined as:
\begin{equation}
\text{Bpref} = \frac{1}{R} \sum_{r=1}^R 1 - \frac{|n \text{ ranked higher than }r|}{R}
\end{equation}

Where $R$ is the number of judged relevant documents, $r$ is a relevant retrieved document, $n$ is one of the first $R$ irrelevant retrieved documents, and non-judged documents are ignored.

\subsection{Results}
As of the submission of this work, two rounds of the TREC-COVID competition have elapsed, and relevance judgements have been generated for each. 
Our results on this data are shown in Table \ref{tab:trec}.
We present our system in two contexts. 
The first is within the general set of submissions. 
This includes metric evaluations on all documents - annotated and non-annotated - and this includes ranking against manual, automatic and feedback systems. 
Manual submissions use human operators to adjust the query or the retrieved documents to improve ranking. 
Feedback systems use supervision from the relevance judgements of prior rounds.
Automated search engines may not do either.
In the second context, we evaluate our system (and all others) strictly on relevance judgements, and we compare our automated system strictly against other automated systems.
To determine rankings, we account for multiple submissions with the same score, and assign to each the highest one (i.e., if the top two scoring submissions for a metric have the same score, each would be ranked \#1). 

In the first context, our system ranks in the top 21 (Round 1, 144 systems) and in the top 3 (Round 2, 136 systems). See Table \ref{tab:trec}. The column Judged@n shows the average percentage, across queries, of topic-documents that have been annotated. As expected, as the percentage rises from Round 1 to Round 2, so do the scores and rankings of the system. In the second context, our system ranks \#1 across metrics in both rounds, against 102 systems in Round 1, and 73 systems in Round 2.

Round 1 used 30 topics and 51,000 documents, yielding 1.53 million possible topic-document pairs. 
There were 8,691 relevance judgements, representing 0.57\% annotation coverage. Round 2 used 35 topics (30 from Round 1), with 63,000 documents, yielding 2.2 million possible topic-document pairs.
There were 12,037 relevance judgements, representing 0.54\% annotation coverage.

\begin{table}[t]
    \caption{TREC-COVID results}
    \centering
    \begin{tabular}{l >{\centering}m{2cm} c >{\centering}m{2cm} >{\centering}m{2cm} r}
        \toprule
        & Score     & Rank  & Score     & Rank & Judged@n \\
        \midrule
        \textbf{Round 1} & \multicolumn{2}{c}{\makecell{Automated Systems (102) \\ Judged Pairs (8,691)}} & \multicolumn{2}{c}{\makecell{All Systems (144) \\ All Pairs (1.53M)}} \\
        \midrule
         Bpref      & 0.5176    & 1      & 0.5176   & 3     & - \\
         MAP        & 0.4870     & 1      & 0.2401   & 21    & - \\
         P@5        & 0.8267    & 1      & 0.6333   & 17    & 72.00\% \\
         P@10       & 0.7933    & 1      & 0.5567   & 21    & 65.67\% \\
         nDCG@10    & 0.7233    & 1      & 0.5445   & 21    & 65.67\%\\
        \toprule
        \textbf{Round 2} & \multicolumn{2}{c}{\makecell{Automated Systems (73) \\ Judged Pairs (12,037)}} & \multicolumn{2}{c}{\makecell{All Systems (136) \\ All Pairs (2.2M)}} \\
        \midrule
         Bpref      & 0.5232    & 1     & 0.5402    & 2     & - \\
         MAP        & 0.5138    & 1     & 0.3487    & 1     & -\\
         P@5        & 0.8171    & 1     & 0.8000    & 3     & 98.29\% \\
         P@10       & 0.7629    & 1     & 0.7200    & 3     & 93.71\% \\
         nDCG@10    & 0.7247    & 1     & 0.6996   & 1     & 93.71\% \\
         \bottomrule
    \end{tabular}
    \label{tab:trec}
\end{table}





\section{Discussion}

Here we present CO-Search, a COVID-19 scientific search engine over the growing corpus of coronavirus literature. 
We train the system using the scientific papers of the COVID-19 Open Research Dataset challenge, and evaluate it against other search engines using the data of the TREC-COVID competition.
Our systems ranks best amongst automated systems, and near the best amongst all systems, as judged using various metrics including nDCG@10, P@5, P@10, MAP, and Bpref.

The system uses a combination semantic and keyword retriever that combines SBERT-derived embeddings with TFIDF \& BM25 vectors to retrieve documents. 
It leverages a Wikipedia \& PubMed pre-trained multi-hop question answering system, together with an abstractive summarizer, to modulate retrieval scores. 
This structure allows CO-Search to disambiguate between subtle word orderings that, in biological contexts, result in critically different meanings (e.g. “What regulates expression of the ACE2 protein?” vs. “What does the ACE2 protein regulate?”), maximizing its utility to the medical and scientific communities in a time of crisis. 



%

\section*{Broader Impact}


This work is intended as a tool to support the fight against COVID-19. In this time of crisis, tens of thousands of documents are being published, only some of which are scientific, rigorous and peer-reviewed. This may lead to the inclusion of misinformation and the potential rapid spread of scientifically disprovable or otherwise false research and data. People on the front lines - medical practitioners, policy makers, etc. - are time-constrained in their ability to parse this corpus, which could impede their ability to approach the returned search results with the appropriate levels of skepticism and inquiry available in less exigent circumstances. Coronavirus-specialized search capabilities are key to making this wealth of knowledge both useful and actionable. The risks are not trivial, as decisions made based on returned, incorrect, or demonstrably false results might jeopardize trust or public health and safety. The authors acknowledge these risks, but believe that the overall benefits to researchers and to the broader COVID-19 research agenda outweigh the risks.



\bibliographystyle{plain}
\bibliography{co-search.bib}



%
%
%

\end{document}